\documentclass[11pt]{article}

\usepackage{amsmath,amssymb}
\usepackage{graphicx}
\usepackage{geometry}
\usepackage{setspace}
\usepackage{hyperref}
\usepackage{caption}
\usepackage{booktabs}

\title{Universal Basic Income with Time-Decaying Currency:\\
Structural Effects on Essential Labor and Long-Term Formation}

\author{
Hitoshi Yamada\\
Toyota Motor Corporation\\
Toyota, Aichi, Japan
}

\date{}

\begin{document}
\maketitle

\begin{abstract}
    Time-decaying currencies have long been discussed in economic theory as a means to discourage hoarding and promote circulation.
    However, their modern digital implementation as a universal basic income (UBI) mechanism raises unresolved structural questions regarding labor participation and long-term social reproduction.
    In this study, we analyze a dual-currency model in which a time-decaying currency is distributed exclusively as UBI, while labor income and savings are denominated in a standard currency.
    Through agent-based simulations, we identify the acceptance ratio of the time-decaying currency for necessities as a critical design parameter.
    Our results show that essential labor does not necessarily collapse under such a system.
    Nevertheless, beyond a threshold acceptance ratio, delayed labor participation and weakened human capital formation emerge even in the absence of material deprivation.
    These findings suggest that time-decaying currency can stabilize short-term living conditions while distorting long-term formation incentives, depending on system design.
\end{abstract}

\section{Introduction}

Universal Basic Income (UBI) has attracted increasing attention as a potential response to structural 
changes in labor markets driven by automation, demographic shifts, and economic volatility.
A wide range of policy experiments and pilot programs have investigated the effects of unconditional income 
transfers on labor supply, consumption stability, and subjective well-being. Empirical studies from developing
 and developed economies alike have generally reported limited or no negative impact on labor participation,
  alongside improvements in consumption smoothing and financial security.

Notable examples include unconditional cash transfer pilots conducted in India, 
which documented increases in labor participation, improvements in health and education outcomes, 
and enhanced household savings capacity.\cite{Banerjee2019} Similar findings have been reported in broader reviews
 of UBI-related interventions, suggesting that unconditional income provision does not necessarily 
 induce large-scale withdrawal from productive activities.\cite{Standing2013India} These results have contributed to a 
 growing consensus that UBI, when evaluated through conventional income transfer mechanisms, 
 does not inherently erode work incentives.

At the same time, recent years have seen growing interest in alternative monetary designs 
intended to complement or extend UBI schemes. Among these, time-decaying or demurrage-based currencies
 have been discussed as mechanisms to encourage circulation, prevent hoarding, and stabilize short-term
  consumption. Historically, the conceptual foundations of such currencies can be traced to early twentieth-century 
  monetary thought, while contemporary discussions increasingly focus on digitally implemented systems 
  enabled by cryptographic technologies.\cite{Gesell1916}\cite{Gelleri2009}

Despite this renewed interest, existing empirical and theoretical studies largely focus on unconditional 
income as a quantity of transfer, rather than on the monetary form through which such income is delivered. 
In particular, little attention has been paid to how design-specific properties of time-decaying currency—such 
as usage constraints and institutional acceptance rules—may introduce dynamic distortions that are not 
captured in standard labor-supply analyses.

Importantly, empirical findings from UBI pilots are inherently context-dependent 
and do not directly address design-dependent distortions induced by time-decaying 
currency mechanisms. While conventional cash-based UBI experiments provide valuable 
evidence regarding income effects, they offer limited insight into how alternative 
monetary architectures may reshape intertemporal incentives, especially when short-term 
consumption support and long-term value storage are institutionally separated.

This study addresses this gap by examining a dual-currency UBI system in which a time-decaying 
currency is allocated unconditionally to cover daily necessities, while a standard non-decaying 
currency is required for savings and long-term accumulation. Using an agent-based model grounded 
in procedural utility maximization under budget and necessity constraints, we analyze how institutional 
parameters governing currency acceptance shape labor participation, savings behavior, and human capital formation dynamics.

Rather than evaluating whether UBI itself is desirable, the present work focuses on how monetary 
design choices condition behavioral outcomes over time. By systematically exploring the parameter 
space of currency acceptance and benefit levels, we identify regimes in which short-term material 
stability coexists with delayed labor participation and weakened formation incentives. 
These results highlight the importance of monetary form as a structural dimension of UBI policy design.\footnote{
Related ideas on digitally implemented time-decaying currency as a UBI mechanism
have been discussed in non-academic contexts in recent years.
}

\section{Model}

\subsection{Dual-currency framework}

We consider a simplified society in which two currencies coexist.

The first is a standard currency, denoted by $Y$, representing conventional fiat money.
All labor income is paid exclusively in $Y$, and only $Y$ can be accumulated as savings or used for future-oriented purposes.
The second is a time-decaying currency, denoted by $D$, which is distributed unconditionally by the government as a universal basic income.
Currency $D$ experiences deterministic value decay and cannot be converted into $Y$.

This separation reflects a central design assumption of the model:
the time-decaying currency functions strictly as short-term consumption support and is not intended to substitute for wages or long-term value storage.
By construction, future-oriented economic capacity must be generated through participation in labor compensated in $Y$.

\subsection{Necessities and acceptance ratio}

Each agent faces a fixed level of monthly necessities.
These necessities are divided into two categories.

The first category consists of necessities that can be paid using either currency $D$ or $Y$.
The second category consists of necessities that must be paid exclusively in $Y$, such as housing costs, medical expenses, or education-related payments.

We define the acceptance ratio $\phi \in [0,1]$ as the fraction of total necessities that belong to the first category.
Thus, $\phi$ represents the degree to which the time-decaying currency is institutionally accepted in daily life.

Importantly, $\phi$ is treated as an exogenous policy parameter.
While the UBI amount determines how much support is provided, $\phi$ determines where and how that support can be used.
As shown later, this distinction plays a central role in shaping system behavior.

\subsection{Labor options}

At each discrete time step, agents choose one of three mutually exclusive actions:
essential labor ($E$), non-essential labor ($N$), or non-work ($0$).

Essential labor represents socially necessary work characterized by higher disutility and fixed aggregate demand.
Non-essential labor yields lower income and lower disutility.
Non-work yields no labor income but allows agents to rely on UBI for consumption.

Agents are heterogeneous with respect to labor disutility.
This heterogeneity is captured by an individual-specific parameter that scales the perceived cost of labor across all actions.

\subsection{Implicit utility structure}

Although the simulation code evaluates agent decisions procedurally, behavior is fully consistent with utility maximization under constraints.

For each agent $i$ and action $a \in \{E, N, 0\}$, we define an implicit utility function:
\[
U_i(a)
=
u\!\left(C_i^D(a)\right)
+
\beta\, u\!\left(S_i^Y(a)\right)
-
\alpha_i\, L(a)
-
\pi\, \mathbf{1}[\text{unmet}(a)].
\]

Here, $C_i^D(a)$ denotes the quantity of necessities actually satisfied using the time-decaying currency $D$, supplemented by $Y$ if required.
$S_i^Y(a)$ represents the remaining amount of standard currency $Y$ after necessities are paid and corresponds to potential savings or future capacity.
$L(a)$ denotes the labor disutility associated with action $a$, satisfying $L_E > L_N > L_0 = 0$.
The parameter $\alpha_i$ captures individual heterogeneity in labor burden.
The indicator function $\mathbf{1}[\text{unmet}(a)]$ equals one if necessities are not fully satisfied under action $a$, in which case a fixed penalty $\pi$ is imposed.
The function $u(\cdot)$ is monotonically increasing and concave; in implementation, it is instantiated as a square-root function.

Agents select actions by maximizing $U_i(a)$.
In the simulation, this maximization is implemented procedurally by explicitly computing and comparing utilities for all actions.
Conditional branching in the code is therefore mathematically equivalent to an argmax operator and does not represent heuristic or rule-based behavior.

\subsection{Modeling intent and scope}

The model intentionally excludes intrinsic motivation, social norms, and non-economic satisfaction derived from labor.
Agents are modeled as economically rational actors seeking to maximize immediate consumption stability and future-oriented financial capacity.

This simplification is deliberate.
By abstracting away from psychological and normative factors, the model isolates distortions arising purely from institutional design.
In particular, it allows us to identify how the interaction between time-decaying currency and necessity coverage shapes incentives independently of behavioral assumptions.

\subsection{Procedural decision-making and equivalence to utility maximization}

Although agent decisions are implemented using explicit conditional branching,
this procedure is mathematically equivalent to utility maximization.

In the code, utilities $U_E$, $U_N$, and $U_0$ are computed for all actions,
and the selected action corresponds to
\[
\arg\max \{ U_E, U_N, U_0 \}.
\]

Conditional statements such as \texttt{if--elif--else} therefore represent
an explicit expansion of the argmax operator rather than rule-based behavior.

\subsection{Separation of simulation and visualization}

The simulation code implementing agent decision logic is fully separated
from visualization and phase-diagram generation scripts.

Visualization modules import simulation outputs but do not influence
agent behavior, ensuring that all reported results share an identical
decision-making structure.

\subsection{Clarification on modeling interpretation}

The model does not replace preferences with heuristic rules.
Instead, utility maximization is implemented procedurally under
non-linear constraints induced by dual currencies and necessity requirements.

This clarification is intended to prevent misinterpretation of the model
as rule-based or non-economic.

\section{Simulation Setup}

We conduct agent-based simulations over a finite time horizon.
Each simulation consists of a population of heterogeneous agents repeatedly making labor participation decisions under fixed institutional parameters.

Key parameters explored in this study include:
the UBI amount distributed in time-decaying currency $B_D$,
the acceptance ratio $\phi$,
and the decay rate $\lambda$ of currency $D$.
Wage levels for essential and non-essential labor are fixed exogenously.

For each parameter combination, we record time series of labor participation, non-work behavior, and unmet necessities.
To characterize system behavior, we focus on aggregated statistics capturing extreme and average outcomes over time.

\section{Results}

\subsection{Essential labor stability}

\begin{figure}[t]
  \centering
  \includegraphics[width=0.85\linewidth]{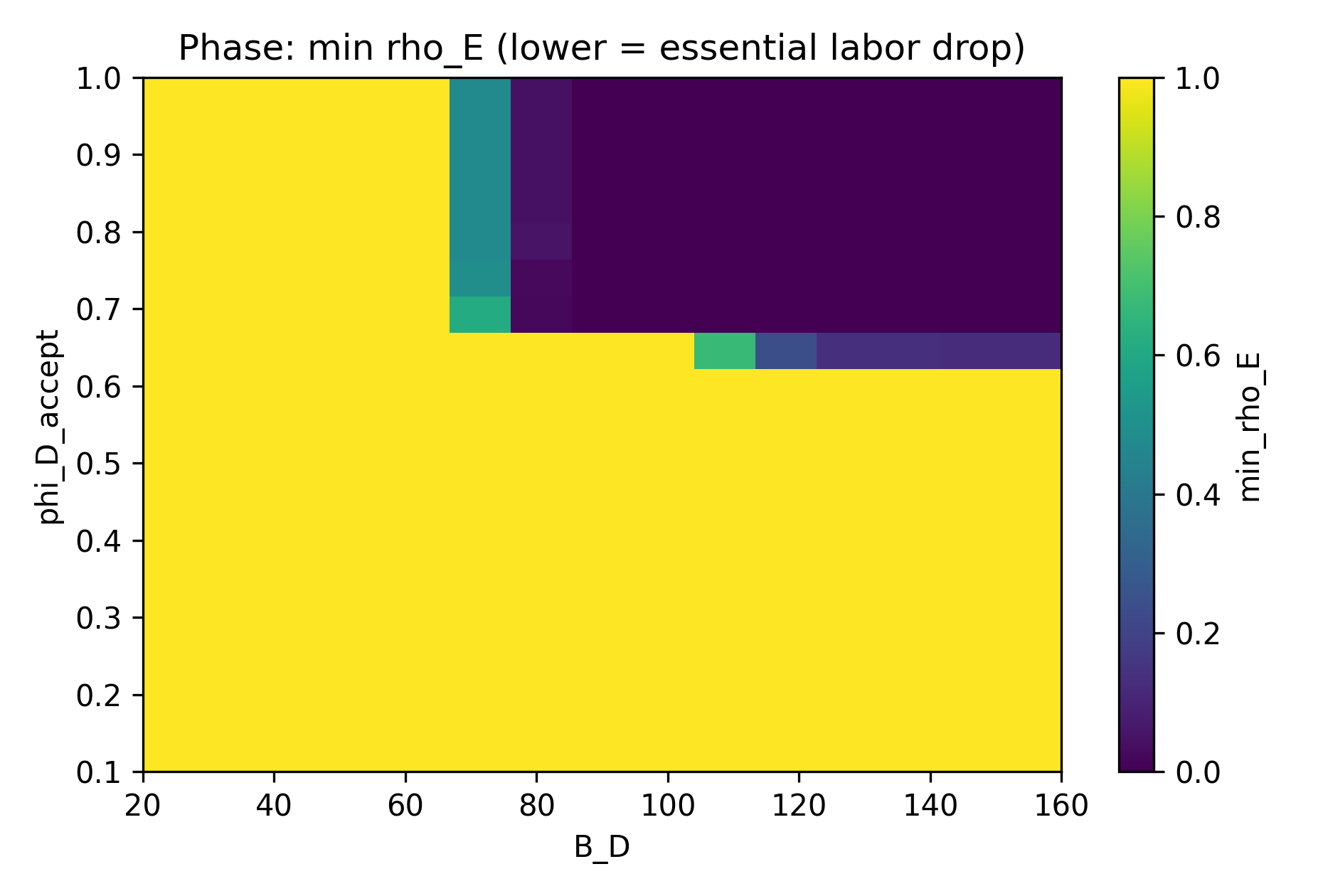}
  \caption{
  Phase diagram of the minimum essential labor supply,
  $\min_t \rho_E(t)$,
  as a function of the UBI benefit level $B_D$ and the acceptance ratio $\phi$.
  Lower values indicate a drop in essential labor participation.
  }
  \label{fig:min_rhoE}
\end{figure}

Figure~\ref{fig:min_rhoE} shows the phase diagram of the minimum essential labor supply,
\[
\min_t \rho_E(t),
\]
where $\rho_E(t)$ denotes the fraction of agents engaged in essential labor at time $t$.

For sufficiently low acceptance ratios $\phi$, essential labor remains stable across the entire range of the UBI benefit level $B_D$ examined in this study.
In this region, the minimum value of $\rho_E(t)$ stays close to unity, indicating that essential services are continuously supplied without disruption.

As $\phi$ increases beyond a critical threshold, a sharp transition is observed.
The minimum essential labor supply drops discontinuously, forming a phase boundary that is largely independent of the benefit level $B_D$.
This indicates that the stability of essential labor is primarily governed by the institutional acceptance condition of the time-decaying currency, rather than by the nominal size of the UBI payment itself.

These results demonstrate that the introduction of a time-decaying currency as UBI does not necessarily lead to the collapse of essential labor.
Instead, essential labor supply exhibits regime-dependent stability determined by the acceptance ratio $\phi$.

---

\subsection{Non-work spikes: instantaneous behavior}

\begin{figure}[t]
  \centering
  \includegraphics[width=0.85\linewidth]{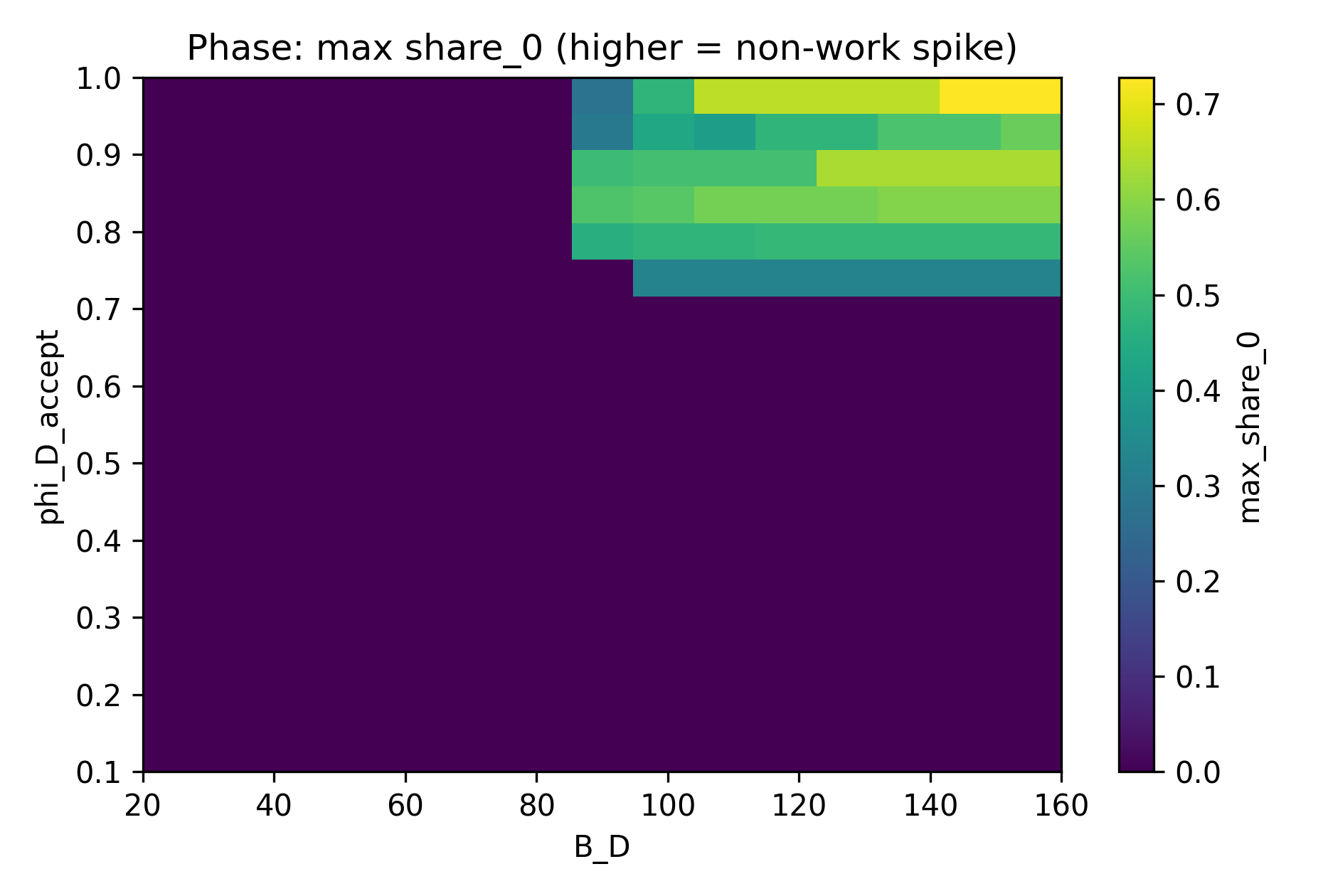}
  \caption{
  Maximum instantaneous share of non-working agents,
  $\max_t \mathrm{share}_0(t)$,
  across the parameter space.
  Higher values correspond to transient collective withdrawals from labor.
  }
  \label{fig:max_share0}
\end{figure}

Figure~\ref{fig:max_share0} presents the maximum instantaneous share of non-working agents,
\[
\max_t \mathrm{share}_0(t),
\]
which captures short-term collective withdrawals from labor.

In regions characterized by high $\phi$ and moderate to high $B_D$, pronounced spikes in non-work behavior emerge.
These spikes occur even in parameter regimes where essential labor remains stable, as shown in Figure~\ref{fig:min_rhoE}.

The observed peaks in $\mathrm{share}_0(t)$ are transient rather than persistent.
They reflect short-term synchronization of non-work decisions across agents, rather than a permanent exit from labor participation.
This distinction is important, as it indicates that non-work spikes do not directly correspond to a breakdown of essential services.

Thus, instantaneous non-work behavior and long-term essential labor stability are not equivalent, and they respond differently to the institutional parameters of the system.

---

\subsection{Average non-work share and formation delay}

\begin{figure}[t]
  \centering
  \includegraphics[width=0.85\linewidth]{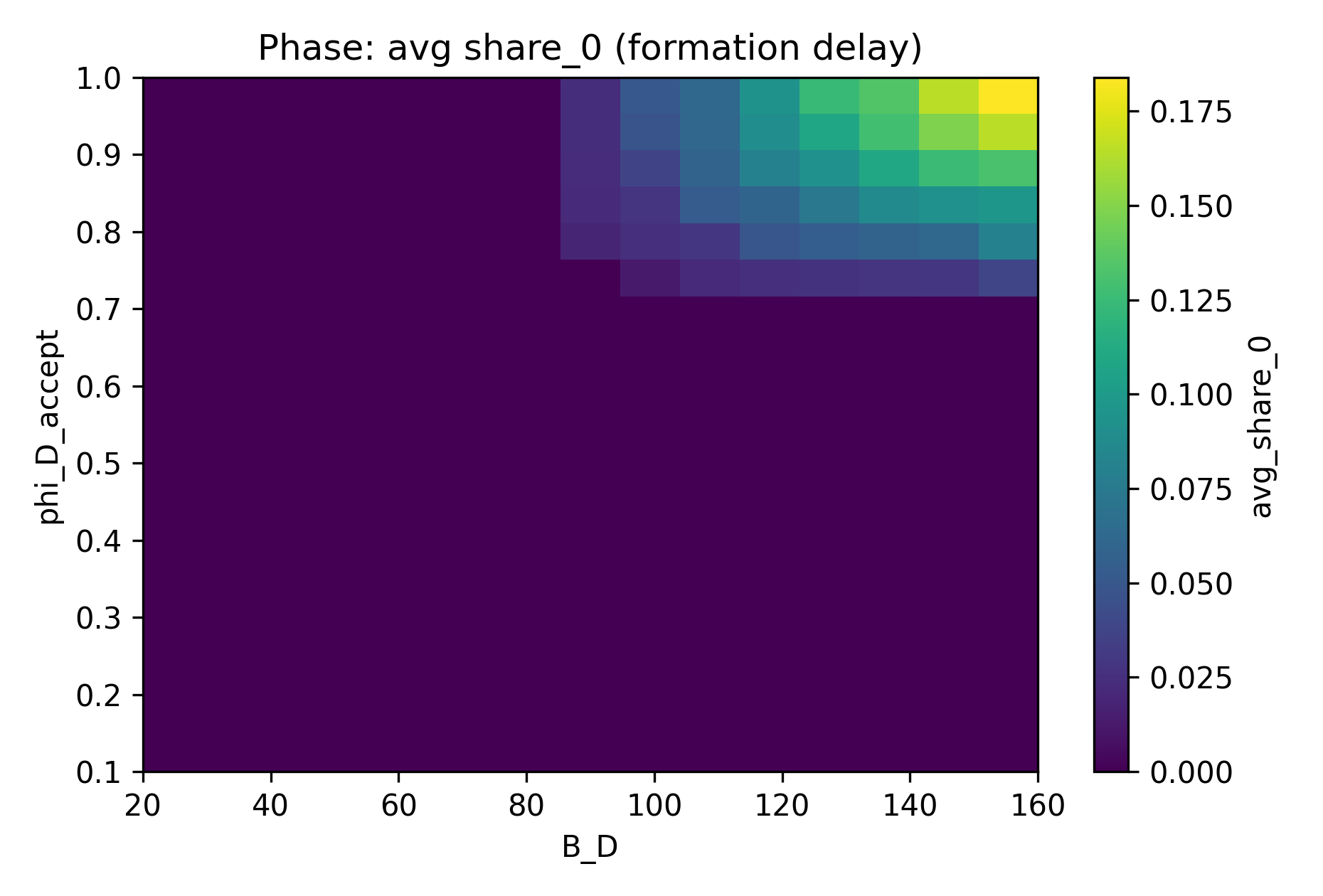}
  \caption{
  Time-averaged share of non-working agents,
  $\langle \mathrm{share}_0(t) \rangle_t$,
  representing typical labor participation over time.
  Higher values indicate delayed labor participation despite stable consumption.
  }
  \label{fig:avg_share0}
\end{figure}

Figure~\ref{fig:avg_share0} shows the time-averaged share of non-working agents,
\[
\langle \mathrm{share}_0(t) \rangle_t,
\]
which represents the typical degree of labor participation over the simulation horizon.

Even in regions where essential labor remains stable and consumption needs are continuously met, the average non-work share increases with higher acceptance ratios $\phi$.
This indicates a systematic delay in labor participation rather than a temporary fluctuation.

Notably, this increase in the average non-work share is not accompanied by material deprivation.
Instead, it reflects a distortion in long-term participation incentives, suggesting a slowdown in formation processes such as human capital accumulation or future-oriented capacity.

The contrast between stable essential labor supply (Figure~\ref{fig:min_rhoE}) and elevated average non-work participation (Figure~\ref{fig:avg_share0}) highlights a structural divergence between short-term system stability and long-term formation dynamics.

---

\subsection{Summary of phase structure}

Taken together, Figures~\ref{fig:min_rhoE}--\ref{fig:avg_share0} reveal a non-trivial phase structure in the dual-currency UBI system.

Essential labor supply can remain stable across a wide range of benefit levels, provided that the acceptance ratio $\phi$ remains below a critical threshold.
However, even within this stable regime, transient non-work spikes and sustained increases in average non-work participation can emerge.

These results indicate that the system does not transition directly from stability to collapse.
Instead, distortions first appear in the form of delayed labor participation and weakened formation dynamics, while essential services continue to be supplied.

\section{Discussion}

The results presented in Figures~\ref{fig:min_rhoE}--\ref{fig:avg_share0} demonstrate that the stability of essential labor supply and individual labor participation incentives are not equivalent outcomes in a dual-currency UBI system with time-decaying money.
While essential services can remain continuously supplied over a wide parameter range, distortions emerge in the temporal structure of labor participation.

A key finding is that the acceptance ratio $\phi$ plays a dominant role in shaping system behavior.
Across all examined benefit levels $B_D$, changes in $\phi$ induce sharp transitions in essential labor stability, whereas variations in $B_D$ alone do not.
This indicates that institutional acceptance conditions---namely, where and how the time-decaying currency can be used---are more consequential than the nominal generosity of the UBI payment.

Importantly, the absence of essential labor collapse does not imply the absence of incentive distortion.
As shown by the coexistence of stable minimum essential labor supply (Figure~\ref{fig:min_rhoE}) and elevated average non-work participation (Figure~\ref{fig:avg_share0}), the system can maintain short-term operational stability while simultaneously weakening long-term participation dynamics.
This divergence highlights a structural limitation of time-decaying money when used as a primary income mechanism.

The time-decaying currency effectively secures short-term consumption by design.
However, because its value erodes over time, it cannot be efficiently accumulated or deferred.
As a result, agents face reduced incentives to engage in forward-looking behaviors that rely on intertemporal planning, such as skill acquisition or sustained labor participation.
The observed increase in time-averaged non-work share is therefore interpreted not as permanent withdrawal from labor, but as a delay in formation processes.

Transient non-work spikes further illustrate this distinction.
Even in regimes where essential labor remains stable, short-term synchronization of non-work behavior emerges under high acceptance ratios (Figure~\ref{fig:max_share0}).
These spikes do not constitute systemic failure but reveal sensitivity in individual decision timing.
They indicate that labor participation responds not only to material sufficiency but also to the temporal properties of the income medium itself.

Taken together, these findings suggest that the primary risk associated with time-decaying currency–based UBI is not immediate labor collapse.
Rather, the more subtle and potentially persistent effect is the erosion of future-oriented capacity, manifested as delayed labor participation and weakened formation dynamics.
Such effects may accumulate gradually and remain obscured if evaluation focuses solely on short-term indicators such as consumption stability or essential service provision.

From a policy design perspective, these results emphasize the importance of acceptance structure over payment size.
Expanding the range of goods and services for which time-decaying currency is accepted can stabilize short-term living conditions, but may simultaneously amplify long-term incentive distortions.
This trade-off is non-trivial and cannot be resolved by adjusting benefit levels alone.

More broadly, the present analysis underscores that UBI systems incorporating non-standard monetary properties should be evaluated along multiple temporal dimensions.
Short-term stability and long-term formation are distinct criteria, and success in one does not guarantee success in the other.
Failure to account for this distinction risks overlooking gradual structural distortions that emerge only over extended time horizons.

\section{Conclusion}

This study examined a dual-currency UBI system in which a time-decaying currency is distributed as basic income, while labor income and savings are denominated in a standard currency.
Using agent-based simulations, we analyzed how institutional parameters, particularly the acceptance ratio $\phi$, shape labor participation dynamics under such a system.

Our results show that the introduction of a time-decaying currency as UBI does not necessarily lead to the collapse of essential labor.
Instead, a more subtle structural effect emerges: while short-term consumption and essential service provision can remain stable, long-term participation dynamics are distorted.
These distortions appear not as permanent labor withdrawal, but as delayed labor participation and weakened formation processes, even in the absence of material deprivation.
Crucially, these effects are governed primarily by acceptance conditions rather than by the nominal size of the UBI payment.

These findings do not constitute a normative judgment against UBI or against the use of time-decaying money per se.
Rather, they highlight a non-trivial trade-off between short-term stability and long-term formation that arises from specific institutional designs.
Future work should explore alternative acceptance structures, hybrid mechanisms, or complementary policies that preserve short-term security while sustaining future-oriented incentives.
More broadly, this study underscores the importance of evaluating UBI systems with non-standard monetary properties along multiple temporal dimensions, beyond immediate stability metrics.

\section*{Acknowledgments}

The author would like to express sincere gratitude to several colleagues and friends
who contributed to the development of this work through discussion and support.

In particular, valuable discussions on time-decaying currency mechanisms and their
structural implications helped clarify the conceptual foundation of this study.
The author thanks Taro Kono, Yuhei Yamaguchi, and Yushi Seki for their insightful
comments and constructive exchanges.

The author is also grateful to Nobuhiko Koga and Takashi Arai for providing a
supportive research environment that made this work possible.

\appendix

\section{Agent Decision Logic and Utility Implementation}

\subsection{Purpose of this appendix}

This appendix clarifies the correspondence between the conceptual utility-based model
described in the main text and its procedural implementation in the simulation code.
Because agent decisions are implemented using explicit conditional branching rather than
symbolic optimization routines, this section demonstrates that the implemented behavior
is mathematically equivalent to constrained utility maximization.

---

\subsection{Implicit utility function}

For each agent $i$ and action $a \in \{E, N, 0\}$, the conceptual utility evaluated at each period is

\[
U_i(a)
=
u\!\left(C_i^D(a)\right)
+
\beta\, u\!\left(S_i^Y(a)\right)
-
\alpha_i\, L(a)
-
\pi\, \mathbf{1}[\mathrm{unmet}(a)].
\]

Here,

\begin{itemize}
  \item $C_i^D(a)$ denotes the amount of necessities actually satisfied using the time-decaying currency $D$
        (supplemented by $Y$ if necessary),
  \item $S_i^Y(a)$ is the remaining balance of the standard currency $Y$, which represents potential savings
        and future-oriented capacity,
  \item $L(a)$ denotes labor disutility, satisfying $L_E > L_N > L_0 = 0$,
  \item $\alpha_i$ captures individual heterogeneity in perceived labor burden,
  \item $\pi$ is a fixed penalty applied when necessities are not fully satisfied,
  \item $u(\cdot)$ is a monotonically increasing concave function (implemented as a square-root function).
\end{itemize}

Agents select actions according to

\[
a_i^* = \arg\max_{a \in \{E,N,0\}} U_i(a).
\]

---

\subsection{Procedural evaluation and equivalence to argmax}

In the simulation code, utilities for all actions are computed explicitly at each period.
The decision rule is implemented using conditional branching of the form:

\begin{verbatim}
if U_E >= max(U_N, U_0):
    choose E
elif U_N >= U_0:
    choose N
else:
    choose 0
\end{verbatim}

This structure is an explicit expansion of the $\arg\max\{U_E, U_N, U_0\}$ operator.
It does not represent rule-based or heuristic behavior.
All decisions are determined solely by utility comparisons under constraints.

---

\subsection{Pseudocode representation}

The agent decision process for a single period can be summarized as follows:

\begin{verbatim}
for each agent i:

    for action a in {E, N, 0}:
        compute labor income in Y
        compute available D and Y balances
        compute satisfied necessities C_D
        compute remaining Y balance S_Y

        if necessities unmet:
            penalty = pi
        else:
            penalty = 0

        U[a] = u(C_D) + beta u(S_Y) - alpha_i L(a) - penalty

    choose a* = argmax_a U[a]
\end{verbatim}

This pseudocode corresponds directly to the implemented simulation logic.

---

\subsection{Budget and necessity constraints}

The use of \texttt{min} and \texttt{max} operators in the code enforces:

\begin{itemize}
  \item budget feasibility under dual currencies,
  \item upper bounds on necessity satisfaction,
  \item non-negativity constraints on consumption and balances.
\end{itemize}

These operators implement economic constraints rather than behavioral rules.

---

\subsection{Separation of simulation and visualization}

The project architecture strictly separates behavioral logic from analysis and visualization.

\begin{itemize}
  \item The simulation module implements all agent decision logic and state transitions.
  \item Visualization scripts import simulation outputs and perform parameter sweeps and aggregation only.
\end{itemize}

Visualization modules do not influence agent behavior.
All reported phase diagrams therefore reflect identical decision-making structures.

---

\subsection{Intentional modeling exclusions}

The model intentionally excludes:

\begin{itemize}
  \item intrinsic motivation or enjoyment of labor,
  \item social norms or moral incentives,
  \item non-economic satisfaction derived from work itself.
\end{itemize}

Agents are therefore treated as economically rational actors maximizing short-term
 consumption stability and future-oriented financial capacity.

This simplification is deliberate and allows structural distortions induced purely by
institutional design to be isolated.

---

\subsection{Reading guidance}

Readers should view the model as a structural analysis rather than a predictive forecast.
 The purpose is not to estimate real-world labor supply levels, but to identify
 design-dependent distortions that arise under pure economic rationality in a dual-currency UBI system.


\begin{thebibliography}{9}

\bibitem{Gesell1916}
Gesell, S. (1916).
\textit{The Natural Economic Order}.
Peter Owen.

\bibitem{Gelleri2009}
Gelleri, C. (2009).
Chiemgauer Regiomoney: Theory and practice of a local currency.
\textit{International Journal of Community Currency Research}, 13, 61--75.

\bibitem{Banerjee2019}
Banerjee, A., Niehaus, P., \& Suri, T. (2019).
Universal basic income in the developing world.
\textit{Annual Review of Economics}, 11, 959--983.

\bibitem{Standing2013India}
G.~Standing,
\textit{Unconditional Basic Income: Two Pilots in Madhya Pradesh},
SEWA--UNICEF Background Note prepared for the Delhi Conference,
May 30--31, 2013.

\end{thebibliography}
\end{document}